\documentclass[twocolumn,preprintnumbers]{revtex4}

\usepackage{graphicx}
\usepackage{epsf}
\usepackage{amsmath,amssymb}
\usepackage{color}
\newcommand{\bvec}{\boldsymbol}

\newcommand{\Mg}{{}^{24}\textrm{Mg}}

\newcommand{\red}[1]{\textcolor{red}{#1}}

\usepackage{ulem}

\begin{document}
\preprint{KUNS-2840, NITEP 80}
\title{Microscopic coupled-channel calculation of proton and alpha inelastic scattering
to the $4^+_1$ and $4^+_2$ states of $^{24}\textrm{Mg}$}

\author{Yoshiko Kanada-En'yo}
\affiliation{Department of Physics, Kyoto University, Kyoto 606-8502, Japan}
\author{Kazuyuki Ogata} 
\affiliation{Research Center for Nuclear Physics (RCNP), Osaka University,
  Ibaraki 567-0047, Japan}
\affiliation{Department of Physics, Osaka City University, Osaka 558-8585,
  Japan}
\affiliation{
Nambu Yoichiro Institute of Theoretical and Experimental Physics (NITEP),
   Osaka City University, Osaka 558-8585, Japan}

\begin{abstract}
\begin{description}
\item[Background]
The triaxial and hexadecapole deformations of the $K^\pi=0^+$ and $K^\pi=2^+$ bands of $^{24}$Mg 
have been investigated by the inelastic scatterings of various probes, including electrons, protons, and alpha($\alpha$) particles,
for a prolonged time. However, it has been challenging to explain the unique properties of the 
scatterings observed for the $4^+_1$ state through reaction 
calculations.
\item[Purpose]
To investigate the structure and transition properties of the $K^\pi=0^+$ and $K^\pi=2^+$ bands of $^{24}$Mg employing the 
microscopic structure and reaction calculations via inelastic proton and $\alpha$-scattering. 
Particularly, the $E4$ transitions to the $4^+_1$ and $4^+_2$ states were reexamined.
\item[Method]
The structure of $^{24}$Mg was calculated employing the variation after the parity and total-angular momentum projections 
in the framework of the antisymmetrized molecular dynamics~(AMD).  
The inelastic proton and $\alpha$ reactions were calculated by the microscopic coupled-channel (MCC) 
approach by folding the Melbourne $g$-matrix $NN$ interaction 
with the AMD densities of $^{24}$Mg.
\item[Results]
Reasonable results were obtained on the properties of the structure, including 
the energy spectra and $E2$ and $E4$ transitions of the $K^\pi=0^+$ and $K^\pi=2^+$ bands 
owing to the enhanced collectivity of triaxial deformation.
The MCC+AMD calculation successfully reproduced the angular distributions of the $4^+_1$ and $4^+_2$ cross sections 
of proton scattering at incident energies of $E_p=40$--100~MeV and $\alpha$-scattering at $E_\alpha=100$--400~MeV.
\item[Conclusions]
This is the first microscopic calculation that described the unique properties of 
the $0^+_1\to 4^+_1$ transition.
In the inelastic scattering to the $4^+_1$ state, 
the dominant two-step process of the $0^+_1\to 2^+_1\to 4^+_1$ transitions and the 
deconstructive interference is the weak one-step process were  essential.
\end{description}
\end{abstract}

\maketitle

\section{Introduction}
The band structure and deformations of $\Mg$ have been investigated by various experimental probes,
including electromagnetic transitions~\cite{Branford:1975ciy,Fifield:1979gfv,Keinonen:1989ltz}
and the inelastic scatterings of 
electrons~\cite{Horikawa:1971oau,Horikawa1972,Nakada:1972,Johnston_1974,Li:1974vj,Zarek:1978cvz,Zarek:1984fm}, 
pions~\cite{Blanpied:1990vd}, 
nucleons~\cite{Haouat:1984zz,Rush:1967zwr,Rush:1968qwc,Swiniarski1969,DeSwiniarski:1973nhi,Zwieglinski:1978zza,Zwieglinski:1978zz,Lombard:1978zz,Kato:1985zz,Horowitz:1969eso,Ray:1979zza,Blanpied:1979im,DeLeo:1979lun,DeLeo:1981zz,Amos:1984aph,Eenmaa:1974jfg}, $^3$He~\cite{Griffiths:1967hrr,VanDerBorg:1979pzv},
 and $\alpha$ particles~\cite{Naqib1968,Rebel:1972nip,VanDerBorg:1979pzv,VanDerBorg:1981qiu,Adachi:2018pql}.
In the positive-parity spectra, the $\{0^+_1,2^+_1,4^+_1\}$ and $\{2^+_2,3^+_1,4^+_2\}$ states 
have been assigned to the $K^\pi=0^+$ ground- and $K^\pi=2^+$ side-bands, respectively.
Regarding these positive-parity bands, the properties of the structure,
including the energy spectra and $E2$ transitions, are described by 
the collective rotation of the prolate~($\beta$) deformation
with the static or vibrational triaxial~($\gamma$) deformation. 
Moreover, the hexadecapole~($\beta_4$) deformations of $^{24}$Mg have since been discussed.

To investigate the $\beta_4$ deformations of $^{24}$Mg, various reaction analyses of
the inelastic electron, pion, proton, and $\alpha$ scatterings to the $4^+_1$ and $4^+_2$ 
states
~\cite{Horikawa1972,Haouat:1984zz,Swiniarski1969,Eenmaa:1974jfg,Lombard:1978zz,Blanpied:1990vd,Ray:1979zza,Blanpied:1979im,DeLeo:1979lun,DeLeo:1981zz,Zwieglinski:1978zza,Rebel:1972nip,VanDerBorg:1979pzv} 
and those of the (quasi)elastic scattering of heavy ions~\cite{Siwek-Wilczynska:1974erz,Gupta:2020}
have been performed employing
collective models.
In the analyses of the inelastic scatterings where the phenomenological fittings of the $\beta_2$ and $\beta_4$ parameters 
with and without the $\gamma$ parameters were performed, the weak (or zero) and remarkable $\beta_4$ deformations
have been attributed to the $K^\pi=0^+$ 
and $K^\pi=2^+$ bands, respectively. 
Regarding the $4^+_2$($K^\pi=2^+$) state, which is strongly populated by inelastic scattering, 
the angular distribution of the cross sections has been described by parameter fitting. 
However, for the $4^+_1$($K^\pi=0^+$) state, such reaction calculations with phenomenological collective models have failed
to describe the angular distribution of the $(p,p')$ cross sections. 
Further, microscopic approaches have been applied to 
calculate the inelastic scattering of protons~\cite{Zwieglinski:1978zz,Amos:1984aph} and the inelastic charge form factors 
\cite{Zarek:1978cvz,Sagawa:1985bwh}; however, no microscopic calculation has 
succeeded in reproducing the property of the $0^+_1\to 4^+_1$ transition. 

In inelastic scattering, the $4^+_1$(4.12~MeV) state is weakly produced 
compared to the $2^+_2$(4.24~MeV) state at close energy.
Hence, the $4^+_1$ cross sections can only be measured with high-quality excitation spectra,
which are sufficient to 
resolve the degenerating $4^+_1$ and  $2^+_2$ states. 
It has been revealed that the $4^+_1$ form factors that were measured by electron scattering displayed 
a strange shape ($q$-dependence), which was different from those of  
the normal $\lambda=4$ transitions and the observed $4^+_2$ form factors \cite{Zarek:1978cvz}.
Furthermore, in the inelastic scattering of protons, a similar strange behavior has been observed in the 
angular distribution of the $4^+_1$ cross sections at incident energies of $E_p \le 65$~MeV, 
which was a challenging to 
explain utilizing the one-step cross sections of the distorted-wave Born approximation (DWBA). 
Detailed analyses have been conducted for proton scattering at $E_p=15$--50~MeV
via coupled-channel~(CC) calculations based on a collective model~\cite{DeLeo:1979lun,DeLeo:1981zz}.
They have suggested the dominant two-step contribution in the $4^+_1$ cross sections 
and also discussed 
effects of the weak direct process of $E4;0^+_1\to 4^+_1$ transition.

Nevertheless, the agreement of the result with the data 
was unsatisfactory, and the inelastic scattering of the proton 
to the $4^+_1$ state is still a puzzle. 
Similarly, in the $\alpha$-scattering at $E_\alpha=$104--120~MeV, 
the multi-step effects proved to be effective in  
the $4^+_1$ cross sections~\cite{Rebel:1972nip,VanDerBorg:1979pzv}. 
These facts indicate that higher-order effects contribute to the inelastic transitions 
of the $4^+_1$ state. 
Moreover, there is no fundamental description of the obtained parameters 
even if the CC calculations with a collective model could qualitatively or quantitatively fit the $4^+_1$ cross sections by 
tuning the adjustable parameters in such phenomenological model analyses, 
and 
serious model ambiguity may arise in  both the structure and reaction parts.
Therefore, a microscopic approach is required to reveal the properties of the $4^+_1$ state via 
inelastic scattering.

In our previous study~\cite{Kanada-Enyo-mg24}, 
we have investigated the proton and $\alpha$ inelastic scattering 
off $\Mg$ via the microscopic coupled-channel~(MCC)
calculation combined with a microscopic structure calculation within the antisymmetrized molecular dynamics (AMD)~\cite{KanadaEnyo:1994kw,KanadaEnyo:1995tb,KanadaEn'yo:1998rf,KanadaEn'yo:2012bj}. 
In the MCC calculations, 
CC reaction calculations are conducted with the 
nucleon-nucleus and $\alpha$-nucleus potentials constructed by folding effective $NN$ interactions with 
diagonal~($\rho$) and 
transition~($\rho^\textrm{tr}$) densities of the target nuclei, which are obtained by microscopic structure models. 
The successful results of the MCC+AMD approach of the 
proton and $\alpha$-scatterings off various $p$- and $sd$-shell nuclei
employing the Melbourne $g$-matrix $NN$ interaction \cite{Amos:2000}, which 
is an effective $NN$ interaction in a nuclear medium
based on a bare $NN$ interaction of the Bonn B potential~\cite{Mac87}, are presented 
in our previous reports~\cite{Kanada-Enyo:2019prr,Kanada-Enyo:2019qbp,Kanada-Enyo:2019uvg,Kanada-Enyo:2020zpl,Kanada-Enyo:2020goh,Ogata:2020umn}.
One of the advantages of utilizing the Melbourne $g$-matrix $NN$ interaction in the MCC approach~\cite{Minomo:2009ds,Toyokawa:2013uua,Egashira:2014zda,Minomo:2016hgc,Minomo:2017hjl} 
is that the interaction exhibits energy and density dependences 
and there is no adjustable parameter in the reaction part owing to its fundamental derivation.
Therefore, the structure calculation can be easily combined with the reaction calculation 
of cross sections
to examine the validity of the adopted structure inputs by comparing the calculated results with
referenced reaction data.
Another advantage of this approach is the fact that the combination of the microscopic structure and reaction calculations can be employed to treat the  
electric transitions and inelastic proton and $\alpha$ scatterings in a unified manner. 

In  previous AMD results obtained from the  $\Mg$ structure calculation,
there were still challenges in precisely reproducing the structure properties; 
the excitation energies were overshot and the transition strengths were underestimated. 
These could indicate that the previous method, AMD with variation 
after parity and total-angular-momentum projections (VAP), 
somewhat underestimates the 
collectivity of the excited states of $\Mg$. Moreover, it failed 
to reproduce the shape of the observed $0^+_1\to 4^+_1$ charge form factors, which was 
a crucial shortcoming in the previous MCC+AMD calculation for the 
reproduction of the $4^+_1$ cross sections 
of proton and $\alpha$-scattering.
In principle, these limitations in $\Mg$ structure calculation could be improved by 
the inclusion of higher-order correlations that are beyond the AMD model space 
with a Slater-determinant. However,  
such an extension of the model space will require huge computational costs. 
Hence, an alternative treatment in the AMD+VAP framework is employed to overcome this 
undershooting of the collectivity in this paper. 
Namely, we apply a version of the AMD+VAP model with fixed nucleon-spins 
instead of optimizing the nucleon-spins in the previous version. 
The fixed-spin version could suitably avoid the undershooting 
problem of the AMD+VAP model and obtain good results for the collectivity 
in the $sd$-shell nuclei, as discussed in 
the structure studies of the shape coexistence phenomena of the nuclei 
around $^{28}$Si~\cite{Kanada-Enyo:2004ere,Kanada-Enyo:2011plo}, and was applied to the 
MCC+AMD calculations of the $p+^{28}\textrm{Si}$ and $\alpha+^{28}\textrm{Si}$ reactions~\cite{Kanada-Enyo:2020zpl}.

Here, we reexamine the transition properties of the positive-parity states of the $K^\pi=0^+$ 
and $K^\pi=2^+$ bands of $\Mg$ via proton and $\alpha$ inelastic scattering 
by the MCC+AMD approach employing improved structure inputs that are
obtained by the fixed-spin version of AMD+VAP. Further, 
we discuss the properties of the $4^+_1$ and $4^+_2$ states 
by comparing the calculated cross sections with the experimental data 
including the recently observed $(\alpha,\alpha')$ data at $E_\alpha=130$ and 386~MeV~\cite{Adachi:2018pql}.

The paper is organized as follows. Sec.~\ref{sec:method} briefly explains 
the AMD framework of $\Mg$ and the MCC approaches for $p+\Mg$ and $\alpha+\Mg$ scatterings.
The AMD results of the structure properties of $\Mg$ are described in Sec.~\ref{sec:results1}.
Next, Sec.~\ref{sec:results2} presents the AMD+MCC results of proton and $\alpha$-scattering 
and discusses the transition properties of the $4^+$ states.
Finally, a summary of the study is presented in Sec.~\ref{sec:summary}. 

\section{Method} \label{sec:method}


In the AMD framework, an $A$-nucleon wave function 
is represented by a Slater determinant of 
single-nucleon Gaussian wave functions as follows:
\begin{eqnarray}
 \Phi_{\rm AMD}({\bvec{Z}}) &=& \frac{1}{\sqrt{A!}} {\cal{A}} \{
  \varphi_1,\varphi_2,...,\varphi_A \},\label{eq:slater}\\
 \varphi_i&=& \phi_{{\bvec{X}}_i}\chi_i\tau_i,\\
 \phi_{{\bvec{X}}_i}({\bvec{r}}_j) & = &  \left(\frac{2\nu}{\pi}\right)^{3/4}
\exp\bigl[-\nu({\bvec{r}}_j-\bvec{X}_i)^2\bigr],
\label{eq:spatial}
\end{eqnarray}
In the equations, ${\cal{A}}$ is the antisymmetrizer and  $\varphi_i$ is
the $i$th single-particle wave function written by the product of the
spatial ($\phi_{{\bvec{X}}_i}$), spin ($\chi_i$), and isospin ($\tau_i$)
wave functions. The nucleon-isospin function, $\tau_i$, is fixed to be proton or neutron. 
In the present version of AMD, we fix the nucleon-spin function, $\chi_i$, to be 
an up-spin $(\chi_{\uparrow})$ or down-spin $(\chi_{\downarrow})$.
The Gaussian centroid parameters, 
${\bvec{Z}}\equiv
\{{\bvec{X}}_1,\ldots, {\bvec{X}}_A\}$, of the single-particle wave functions, 
are assumed to be independent complex parameters, which are
determined by the energy optimization for each $J^\pi$ state of $\Mg$.

The energy variation is performed after the parity and total-angular-momentum projections
to minimize the energy expectation value,
$E=\langle \Psi|{\hat H}|\Psi\rangle /\langle \Psi|\Psi\rangle$, for
$\Psi=P^{J\pi}_{MM'}\Phi_{\rm AMD}({\bvec{Z}})$ that is projected from the AMD wave function. 
Here $P^{J\pi}_{MM'}$ is the parity and total-angular-momentum projection operator
and $M'$ is the quanta of the 
$Z$ component~($J_Z$) of the total-angular-momentum in the body-fixed frame. Note that 
$M'$ is not necessarily equal to the $K$ quanta, which is defined by the principal axis of the intrinsic shape,
because the principal axis could tilt from the $Z$-axis during the energy variation.  
VAP is performed for $J^\pi=0^+$, $2^+$, $3^+$, and $4^+$ 
to obtain the states of the $K^\pi=0^+$ and $K^\pi=2^+$ bands.
We select $M'=0$, 1, 2, and 2 for $J^\pi=0^+$, $2^+$, $3^+$, and $4^+$, respectively, 
so as to obtain the minimum energy   
$E=\langle \Psi|{\hat H}|\Psi\rangle /\langle \Psi|\Psi\rangle$ after the variation 
for a given $J^\pi$.
After VAPs of four sets of $J^\pi$ and $M'$, the resulting
four configurations, $\Phi_{\rm AMD}(\bvec{Z}^{(m)})$ 
($m=1,\ldots,4$), are superposed to calculate the final wave functions of positive-parity states.
Specifically, the diagonalization of the Hamiltonian and norm matrices is performed 
by the basis wave functions, $P^{J\pi}_{MM'}\Phi_{\rm AMD}({\bvec{Z}}^{(m)})$,
regarding $M'$ and $m$, which correspond to the $K$-mixing and 
configuration-mixing, respectively.

In the previous AMD+VAP calculation of $\Mg$,  
the nucleon-spin functions, $\chi_i$, were not fixed; they were optimized by the energy variation.
In this paper, we name the present and previous AMD+VAP calculations, i.e., 
those without and with nucleon-spin optimization, ``AMD(fix-s)'' and ``AMD(opt-s)'', respectively.

The effective nuclear interactions employed in the present structure calculation of AMD(fix-s)
are the same as those used in our papers~\cite{Kanada-Enyo-mg24,KanadaEn'yo:1998rf,Kanada-Enyo:2020zpl,Kanada-Enyo:2020goh}. 
The MV1 (Case 1) central force \cite{TOHSAKI} with the parameters, $(b,h,m)=(0,0,0.62)$,
and the spin-orbit term of the G3RS force \cite{LS1,LS2} with strength parameters,
$u_{ls}\equiv u_{I}=-u_{II}=3000$ MeV, are employed.
Coulomb force is also included.


The elastic and inelastic cross sections of the proton and $\alpha$-particle scatterings off $\Mg$ are calculated by MCC+AMD in the same way as done in the previous work~\cite{Kanada-Enyo-mg24}.
The nucleon-nucleus potentials are constructed 
in a microscopic folding model~(MFM) where the diagonal and coupling potentials are calculated
by folding the Melbourne $g$-matrix $NN$ interaction \cite{Amos:2000} with 
diagonal ($\rho(r)$) and transition ($\rho^\textrm{tr}(r)$) densities of the target nucleus. 
The $\alpha$-nucleus potentials are obtained by an extended nucleon-nucleus
folding (NAF) model \cite{Egashira:2014zda}, which is obtained by folding the calculated
nucleon-nucleus potentials with an $\alpha$ density.
For more details of the reaction calculations, the reader is referred to 
the previous paper~\cite{Kanada-Enyo-mg24} and references therein.

For use in the MFM calculation of the nucleon-nucleus potentials, 
$\rho(r)$ and $\rho^\textrm{tr}(r)$ of $\Mg$ 
are calculated from the wave functions obtained from AMD(fix-s).
The charge symmetry breaking in the wave functions obtained for the $K^\pi=0^+$ and 
$K^\pi=2^+$ bands of $\Mg$ is less than several percentages, and hence, it is omitted in the MFM calculation. 
Namely, the proton and neutron components of the densities are averaged as
$\rho(r)=(\rho_p(r)+ \rho_n(r))/2$ and 
$\rho^\textrm{tr}(r)=(\rho^\textrm{tr}_p(r)+ \rho^\textrm{tr}_n(r))/2$, where
$\rho_p(r)\approx \rho_n(r)$ and $\rho^\textrm{tr}_p(r)\approx \rho^\textrm{tr}_n(r)$. 
The $E2$ and $E4$ transition strengths are calculated by the proton transition densities 
as 
\begin{align}
B(E\lambda;J_i\to J_f)=\frac{e^2}{2J_f+1}\Bigl|\int r^\lambda \rho^\textrm{tr}_p(r) r^2 dr\Bigr|^2.
\end{align}
Similarly, the isoscalar transition strengths are given as 
\begin{align}
B(\textrm{IS}\lambda;J_i\to J_f)/4&=\frac{e^2}{2J_f+1}\Bigl|\int r^\lambda \rho^\textrm{tr}(r) r^2 dr\Bigr|^2
\nonumber\\
&\approx B(E\lambda;J_i\to J_f).
\end{align}

To reduce the model ambiguity of the structure calculation, 
the calculated $\rho^\textrm{tr}(r)$
are renormalized by the factor, $f^\textrm{tr}$
as $\rho^\textrm{tr}(r)\to f^\textrm{tr}\rho^\textrm{tr}(r)$
to fit the experimental electromagnetic transition strengths and charge form factors.
For excitation energies of $\Mg$, the experimental values are utilized.

\section{Structure of $^{24}$Mg} \label{sec:results1}

In this section, we present the AMD(fix-s) results of the structure calculation of $\Mg$ and 
compare them to the previous AMD(opt-s) results. 

First we analyze each AMD configurations, 
which are obtained by VAP for the $0^+_1(K=0)$, $2^+_1(K=0)$, $4^+_1(K=0)$, and $3^+_1(K=2)$ states, 
before the superposition
to discuss the intrinsic shapes of these states.
In Table ~\ref{tab:defo}, we present the 
deformation parameters, $\beta$ and $\gamma$, obtained from the expectation values, 
$\langle X^2 \rangle$, $\langle Y^2 \rangle$, and $\langle Z^2 \rangle$, of the 
intrinsic wave functions without the parity and total-angular-momentum projections. 
The result of the AMD(fix-s) calculation is compared to that of 
the AMD(opt-s) calculation.
The AMD(fix-s) calculation obtains larger $\gamma$ values than the previous calculation does,
implying that 
the triaxial collectivity is enhanced in the AMD(fix-s) result, in which 
the $\beta$ and $\gamma$ values are almost constant in the $K^\pi=0^+$ ground-band.

\begin{figure}[!h]
\includegraphics[width=9 cm]{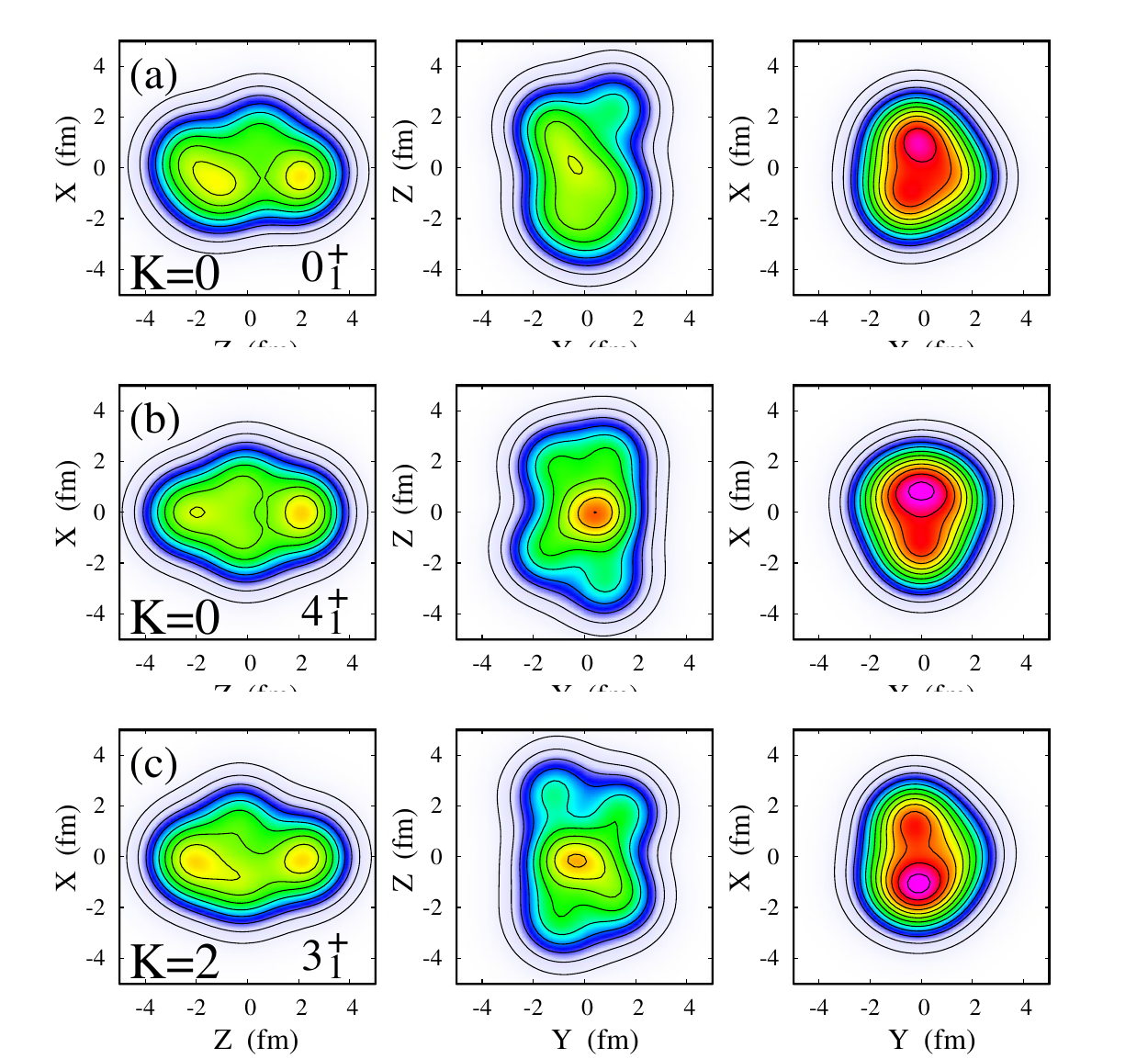}
  \caption{Density distribution of the intrinsic wave functions before the parity and total-angular-momentum projections of the $0^+_1$($K^\pi=0^+$),
$4^+_1$($K^\pi=0^+$), and $3^+_1$($K^\pi=2^+$) states obtained by VAP.
In the left, middle, and right panels,
the densities are projected onto the $X$-$Z$, 
$Y$-$Z$, and $Y$-$X$ planes by integrating along the $Y$, $X$, and $Z$ axes, respectively. 
The intrinsic axes are selected as the principal axes in the following order:
$\langle ZZ\rangle\ge \langle YY\rangle\ge \langle XX\rangle$.
  \label{fig:dense-cont}}
\end{figure}

The intrinsic density distribution of the AMD(fix-s) result 
is displayed in Fig.~\ref{fig:dense-cont}. From the density distribution, 
the intrinsic states exhibit higher-order correlations beyond the quadrupole deformations,
including the cluster components. Moreover, \red{the intrinsic} structure change occurs
with an increase in $J$ along the $K^\pi=0^+$ ground-band
even though the quadrupole deformations are almost unchanged.

\begin{table}[ht]
\caption{Intrinsic deformations, $\beta$ and $\gamma$ (degree), of the 
$0^+_1(K=0)$, $2^+_1(K=0)$, $4^+_1(K=0)$, and $3^+_1(K=0)$ states of $\Mg$ obtained by VAP.
The deformation parameters are calculated for the intrinsic wave functions of single configurations
before the projections and configuration mixing.
The AMD(fix-s) result is compared to the AMD(opt-s) one. 
 \label{tab:defo}
}
\begin{center}
\begin{tabular}{llrrrrrrrrrrrrccccc}
\hline
\hline
$J^\pi_i$~(band)	&	AMD(fix-s) & {AMD(opt-s)}	\\
$0^+_1(K=0)$	&$(	0.35 	,	10 	)$ & $(	0.35 	,	3 	)$\\
$2^+_1(K=0)$	&$(	0.37 	,	12 	)$ & $(	0.33 	,	6 	)$\\
$4^+_1(K=0)$	&$(	0.37 	,	12 	)$ & $(	0.29 	,	7 	)$\\
$3^+_1(K=2)$	&$(	0.37 	,	20 	)$ & $(	0.34 	,	12 	)$\\
\hline
\hline
\end{tabular}
\end{center}
\end{table}

\begin{figure}[!h]
\includegraphics[width=8.5 cm]{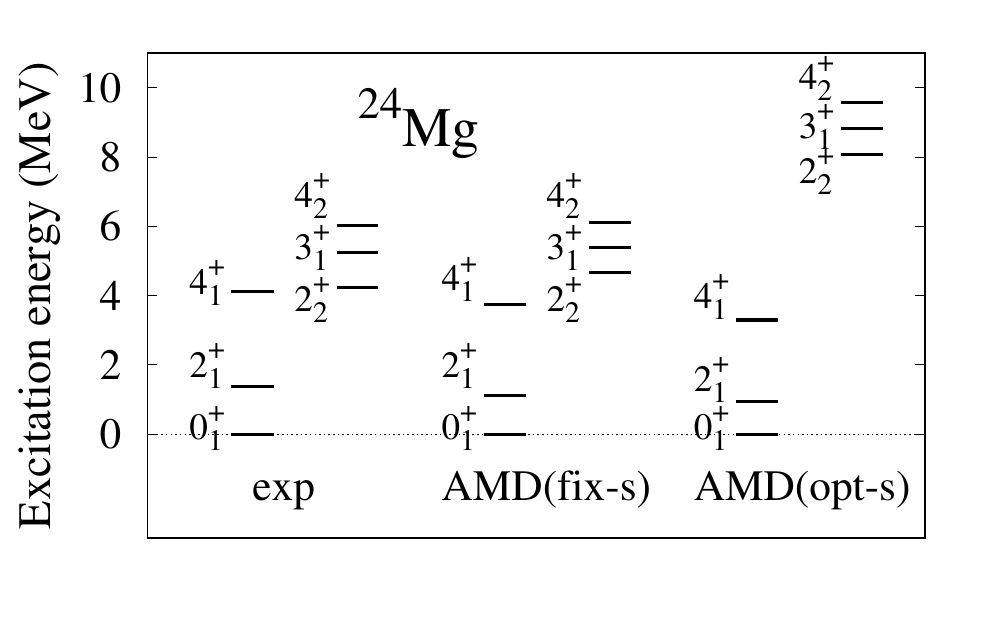}
  \caption{Energy spectra of $\Mg$ for the $K^\pi=0^+$ ground- and $K^\pi=2^+$ side-bands. 
The calculated spectra of the AMD(fix-s) and AMD(opt-s) results 
are compared to the experimental spectra~\cite{Firestone:2007crk}. 
  \label{fig:spe}}
\end{figure}

\begin{table*}[ht]
\caption{$E2$ transition strengths and the electric $Q$ moment of $\Mg$.
The experimental data are from 
Refs.~\cite{Firestone:2007crk,Keinonen:1989ltz,Branford:1975ciy}.
The $B_\textrm{th}$ values obtained by AMD(fix-s) are listed 
with the renormalization factors, $f^\textrm{tr}$, employed for the reaction 
calculations.
For comparison,
the AMD(opt-s) results are also listed. 
The units of the $E2$ transition strengths 
and $Q$ moments are $e^2$fm$^{4}$ and $e$fm$^{2}$, respectively. 
 \label{tab:BE2}
}
\begin{center}
\begin{tabular}{ll|rrr|r|rrrrrrrrccccc}
\hline
\hline
$J^\pi_i$~(band)	&	$J^\pi_f$~(band) &{exp}~\cite{Firestone:2007crk} & exp~\cite{Branford:1975ciy} &
exp~\cite{Keinonen:1989ltz}	&	AMD(opt-s) &	\multicolumn{3}{c}{AMD(fix-s)}	& \\
	&		&	$B(E2)$& $B(E2)$
& $B(E2)$ 	& $B_\textrm{th}(E2)$
&	$B_\textrm{th}(E2)$	& 	$f^\textrm{tr}$(set-I)& $f^\textrm{tr}$(set-II)\\
$2^+_1$ $(K=0)$	&	$0^+_1$ $(K=0)$	&	88.4(4.1)	&	84.3(2.5)	&	86.5(1.6)	&	55.4 	&	70.2 	&	1.12 	&	1.12 	\\
$4^+_1$ $(K=0)$	&	$2^+_1$ $(K=0)$	&	160(16)	&	95(16)	&	155(12)	&	72.8 	&	101.7 	&	1.26 	&	1.11\footnote{Fitting the mean value of Refs.~\cite{Branford:1975ciy} 
and Ref.~\cite{Keinonen:1989ltz}.} 	\\
	&		&		&		&		&		&		&		&		\\
$3^+_1$ $(K=2)$	&	$2^+_2$ $(K=2)$	&	240(30)	&	140(25)	&	157(23)	&	103.1 	&	120.3 	&		&		\\
$4^+_2$ $(K=2)$	&	$2^+_2$ $(K=2)$	&	77(10)	&	66(12)	&	76.9(9.9)	&	36.2 	&	41.2 	&	1.37 	&	1.37 	\\
$4^+_2$ $(K=2)$	&	$3^+_1$ $(K=2)$	&		&		&		&	73.3 	&	81.7 	&		&		\\
	&		&		&		&		&		&		&		&		\\
$2^+_2$ $(K=2)$	&	$0^+_1$ $(K=0)$	&	8.0(0.8)	&	5.8(1.2)	&	6.5(0.5)	&	2.1 	&	3.2 	&	1.58 	&	1.58 	\\
$2^+_2$ $(K=2)$	&	$2^+_1$ $(K=0)$	&	12.2(0.9)	&	11.1(1.6)	&	12.2(0.9)	&	0.6 	&	4.1 	&	1.72 	&	1.72 	\\
$3^+_1$ $(K=2)$	&	$2^+_1$ $(K=0)$	&	10.3(1.2)	&	8.6(1.2)	&	9.6(0.9)	&	3.1 	&	5.6 	&		&		\\
$4^+_2$ $(K=2)$	&	$2^+_1$ $(K=0)$	&	4.1(0.4)	&		&	4.1(0.4)	&	2.9 	&	0.1 	&	1\footnote{no renormalization.} 	&	1$^b$	\\
	&		&		&		&		&		&		&		&		\\
$J^\pi$~(band)	&&{exp}~\cite{Firestone:2007crk} & &
	&	AMD(opt-s) &	\multicolumn{3}{l}{AM(fix-s)}	\\

	&		&	$Q$	&		&		&	$Q$	&	$Q$	&		&		\\
$2^+_1$ $(K=0)$	&		&	$-16.6(6)$	&		&		&	$-15.1$	&	$-17.0$	&		&		\\
\hline
\hline
\end{tabular}
\end{center}
\end{table*}

\begin{table*}[ht]
\caption{$E2$ and $E4$ transition strengths of the $2^+ \to 0^+_1$ and  $4^+ \to 0^+_1$ transitions.
For the experimental values, 
$B(E\lambda)$ from the $\gamma$-decay data \cite{Firestone:2007crk}, 
$B(C\lambda)$ from the $(e,e')$ data \cite{Zarek:1978cvz,Johnston_1974}, 
$B(\textrm{IS}\lambda)/4$ from the 
$(\alpha,\alpha')$ data \cite{VanDerBorg:1981qiu}, and $B(C\lambda)$ from the $(\pi,\pi')$ 
data~\cite{Blanpied:1990vd} are listed.
Regarding the theoretical values, the original values $B_\textrm{th}(\textrm{IS}\lambda)/4$ before the renormalization and 
the renormalized values $f^2_\textrm{tr}B_\textrm{th}(\textrm{IS}\lambda)/4$
are listed together with the renormalization factors $f^\textrm{tr}$(set-I and -II), which are common for the sets I and II
cases.  
The unit of the transition strengths is $e^2$fm$^{2\lambda}$.
\label{tab:BEL}
}
\begin{center}
\begin{tabular}{l|rrrrr|r|rrrrrrrccccc}
\hline
\hline
	&	$\gamma$-decays~\cite{Firestone:2007crk}	&	$(e,e')$~\cite{Zarek:1978cvz}	&	$(e,e')$~\cite{Johnston_1974}	&	$(\alpha,\alpha')$~\cite{VanDerBorg:1981qiu}	&	$(\pi,\pi')$~\cite{Blanpied:1990vd}		& AMD(opt-s) &	\multicolumn{3}{c}{AMD(fix-s)}\\
$J^\pi$~(band)	&	$B(E\lambda\downarrow)$	&	$B(C\lambda\downarrow)$	&	$B(C\lambda\downarrow)$	&	$B(\textrm{IS}\lambda\downarrow)/4$	&	$B(C\lambda\downarrow)$	
	& $B(\textrm{IS}\lambda\downarrow)/4$ &	\multicolumn{2}{c}{$B(\textrm{IS}\lambda\downarrow)/4$} &	$f^\textrm{tr}$\\
	&	
	&		&		&	
	&		&	original	&	original	&	normalized &		\\
$2^+_1$ $(K=0)$	&	88.4(4.1)	&	90.6(7.0)	&	105(5)	&	84 	&	108	&	54 	&	68 	&	86 	&	1.12	\footnote{$f^\textrm{tr}$ determined to fit the $B(E\lambda)$ data~\cite{Firestone:2007crk}.}	&	\\
$2^+_2$ $(K=2)$	&	8.0(0.8)	&	5.48(0.60)	&	5.26(1.2)	&	14 	&	6.7	&	2.0 	&	3.1 	&	7.7 	&	1.58	$^a$	&	\\
$4^+_1$ $(K=0)$	&		&	200(30)	&		&	1200 	&		&	1.1 	&	243 	&	477 	&	1.4 	\footnote{$f^\textrm{tr}$ determined to fit the charge form factors~\cite{Zarek:1978cvz}.}	&	\\
$4^+_2$ $(K=2)$	&		&	4800(600)	&	4700(1100)	&	4700 	&	2900	&	1740 	&	3350 	&	4050 	&	1.1	$^b$	&	\\
\hline
\end{tabular}
\end{center}
\end{table*}

\begin{figure}[!h]
\includegraphics[width=9 cm]{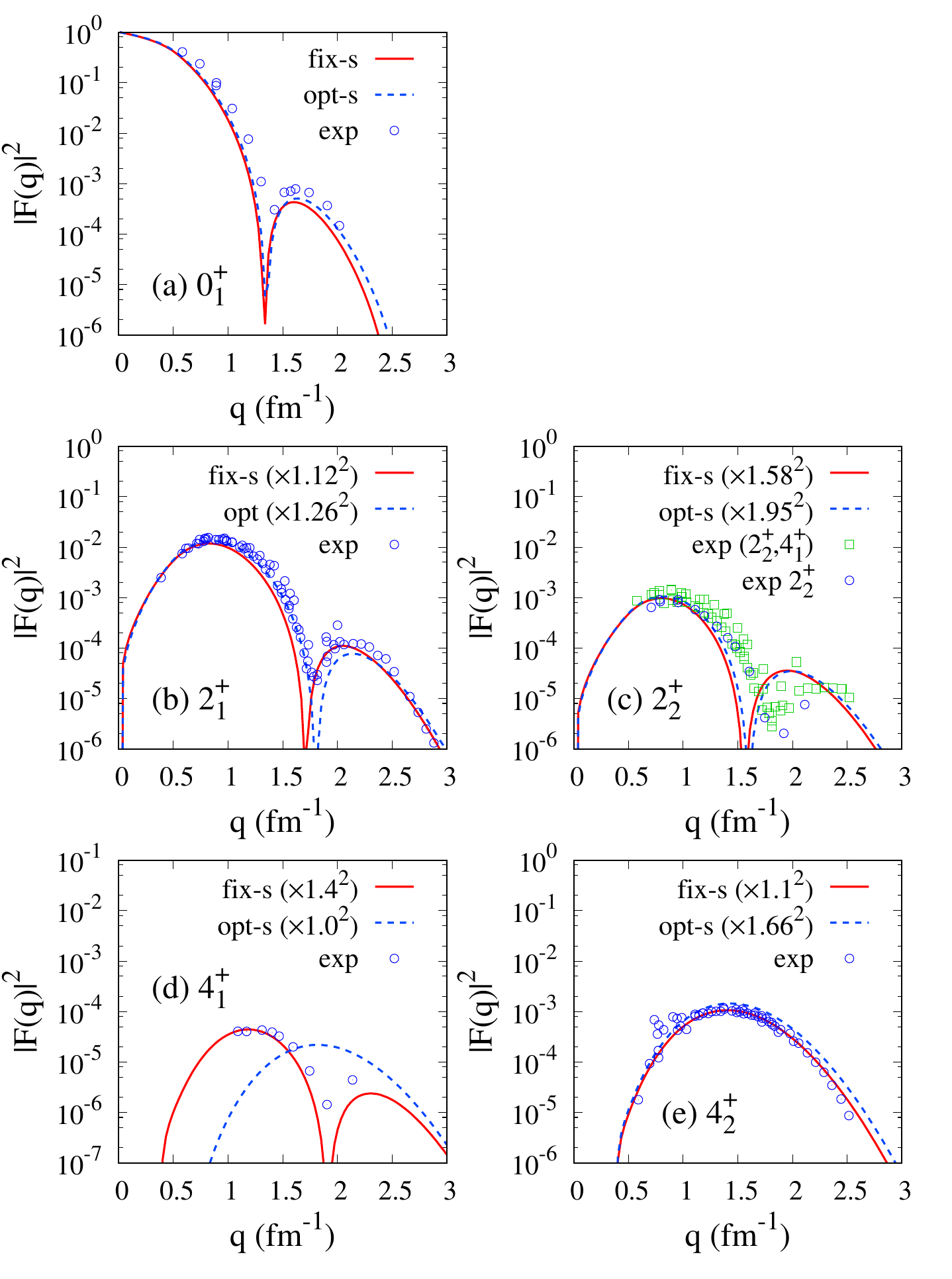}
  \caption{Squared charge form factors of the elastic and inelastic processes of $\Mg$.
For the AMD(fix-s) result, the renormalized form factors
multiplied by the $f^\textrm{tr}$ values in Table~\ref{tab:BEL} are plotted.
For comparison, the AMD(opt-s) result of 
the renormalized form factors are also shown. 
The experimental data are those measured by electron scattering~\cite{Horikawa:1971oau,Nakada:1972,Zarek:1978cvz,Li:1974vj,Johnston_1974}.
In the data for the $2^+_2$(4.24 MeV) state in Refs.~\cite{Horikawa:1971oau,Nakada:1972,Li:1974vj}, 
the $4^+_1$(4.12 MeV) contribution was not separated. 
  \label{fig:form}}
\end{figure}

\begin{figure}[!h]
\includegraphics[width=6 cm]{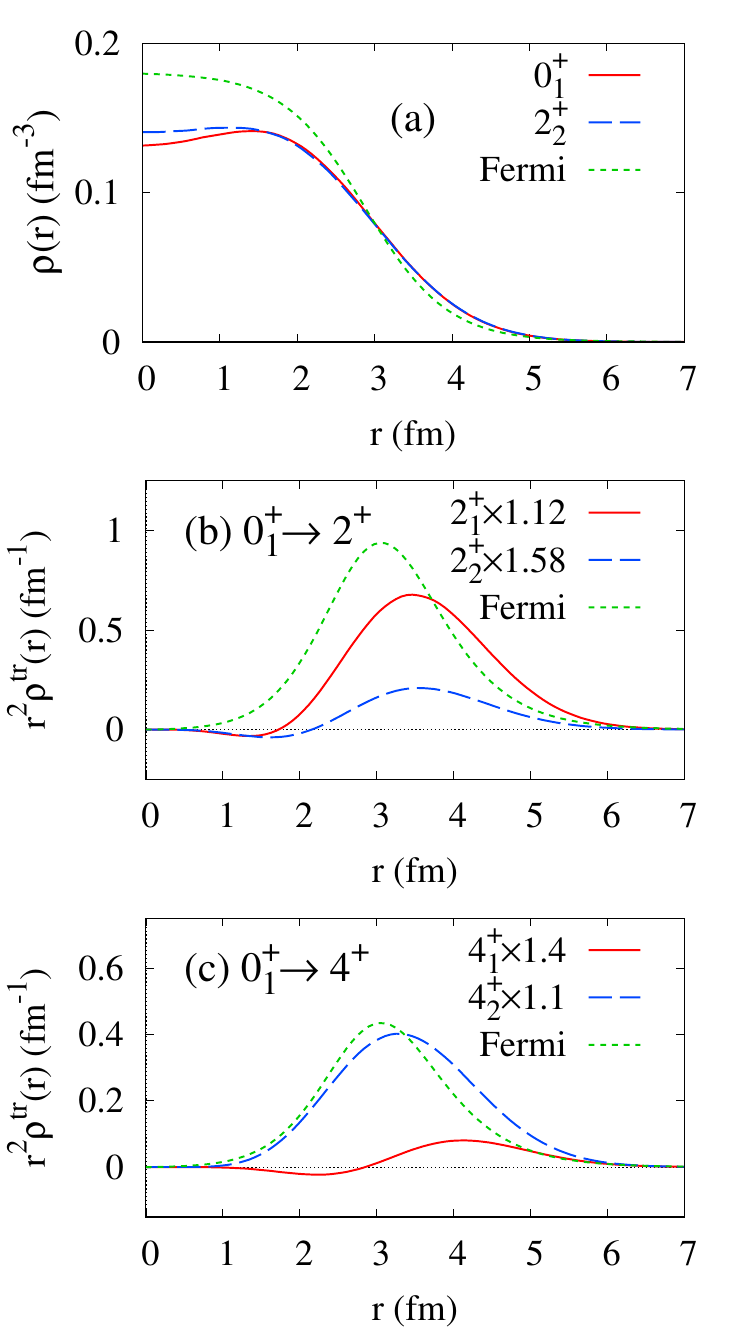}
  \caption{Calculated matter densities and mean values of transition densities 
$\rho^\textrm{tr}(r)=(\rho^\textrm{tr}_p(r)+\rho^\textrm{tr}_n(r))/2$ of the proton and neutron 
components
of $\Mg$ obtained by AMD(fix-s).
The renormalized transition densities, $f^\textrm{tr}\rho^\textrm{tr}_\textrm{th}$,  
employing the factors, $f^\textrm{tr}$, 
listed in Table~\ref{tab:BEL} are shown.
For comparison, the Fermi density $\rho^{}_\textrm{Fermi}(r)=\rho _0 [1+\exp(\frac{r-c}{t/4.4}) ]^{-1}$ 
with $c=2.876$~fm and $t=2.333$~fm, and 
the collective-model transition densities of the Fermi-type Tassie form,
$\rho^\textrm{tr}_\textrm{Tassie}(r)\propto r^{\lambda-1}\partial \rho^{}_\textrm{Fermi}(r)/\partial r$,
are also shown. $\rho^\textrm{tr}_\textrm{Tassie}(r)$ is normalized to fit $B(E2;2^+_1 \to 0^+_1)$ and 
$B(E4;4^+_2 \to 0^+_1)$ of the AMD(fix-s) result.
  \label{fig:trans}}
\end{figure}

Next, we discuss the structure properties obtained by the $K$-mixing and configuration mixing 
after the parity and total-angular-momentum projections.
Figure \ref{fig:spe} shows the comparison between the AMD(fix-s) and  AMD(opt-s) results of 
the energy spectra 
with the experimental data of the ground~($K^\pi=0^+$)- and side~($K^\pi=2^+$)-bands.
In the calculations, the $K^\pi=2^+$ band, including the $2^+_2$, $3^+_1$, and $4^+_2$ states,
is constructed by the 
triaxially deformed intrinsic shape and regarded as the side-band of the $K^\pi=0^+$ ground-band.
Because of the enhanced triaxiality, 
the overshooting shortcoming of the calculated $K^\pi=2^+$ band energy, which was observed
in the AMD(opt-s) result is
sufficiently improved, and a good correlation with the data is obtained in the AMD(fix-s) result. 

The AMD(fix-s) and AMD(opt-s) results of the $E2$ transition strengths, $B_\textrm{th}(E2)$, are 
listed in Table~\ref{tab:BE2} and compared to the experimental data, $(B_\textrm{exp}(E2)$).
The previous AMD(opt-s) calculation underestimated the in-band $E2$ transition strengths by a factor of two, 
but the results are enhanced in the AMD(fix-s) calculation. 
A possible reason for these improvements in the energy spectra and transition strengths of the
AMD(fix-s) calculation
is the following: in the one-Slater description of the AMD+VAP framework, 
the optimization of the nucleon-spins possibly smears the collectivity 
of the deformations in the med-$sd$-shell nuclei, but the 
smearing effect is weakened by fixing the nucleon-spins in the AMD(fix-s) calculation.
Nevertheless, to precisely fit the data of the in-band transition strengths, 
a 10\%--40\% enhancement of the transition matrix elements is still required of the AMD(fix-s) result. 

For use in the MCC calculation, we renormalize the calculated $\rho^\textrm{tr}(r)$ 
with the factors, $f^\textrm{tr}=\sqrt{B_\textrm{exp}(E2)/B_\textrm{th}(E2)}$, to reproduce the experimental transition strengths.
In the default MCC calculation, we utilize the $f^\textrm{tr}$
values that are adjusted the $B_\textrm{exp}(E2)$ values from Ref.~\cite{Firestone:2007crk} in which 
the evaluation was performed with a couple of data sets of lifetime measurements. We call this choice ``$f^\textrm{tr}$(set-I)''. 
In Table~\ref{tab:BE2}, we also present two data sets reported 
in Refs.~\cite{Branford:1975ciy} and \cite{Keinonen:1989ltz}. 
For most of the transitions, 
the two measurements obtained the consistent $B(E2)$ values with each other, but inconsistent values 
for the $4^+_1\to 2^+_1$ transition.
To discuss the effect of this ambiguity to the reaction analysis, 
we adopt another choice of renormalization $f^\textrm{tr}$ for the $4^+_1\to 2^+_1$ transition 
by fitting the mean value of the two data. 
We call this optional choice ``$f^\textrm{tr}$(set-II)''.
The adopted $f^\textrm{tr}$ values of the default case (set-I) and the optional case (set-II)
are listed in Table~\ref{tab:BE2}. The only difference between the two sets, $f^\textrm{tr}$(set-I) and  $f^\textrm{tr}$(set-II),
is the value for the $4^+_1\to 2^+_1$ transition. 

The $E2$ and $E4$ transition strengths of the inelastic transitions from the ground state are listed 
in Table~\ref{tab:BEL}, 
where the calculated values of the isoscalar (IS) components, $B(\textrm{IS}\lambda)/4$, are listed together with 
the experimental values, $B(E\lambda)$, measured by $\gamma$-decays, and  
the empirical values, $B(C\lambda)$, which were evaluated 
by the $(e,e')$ and $(\pi,\pi')$ data, and $B(\textrm{IS}\lambda)/4$ by the $(\alpha,\alpha')$ data. 
The $E4$ transition strengths from the ground state concentrate in the $4^+_2$ state, whereas the transition to the $4^+_1$ state is much weaker.
Compared to the AMD(opt-s) result,  
a significant improvement is achieved for the $4^+_1\to 0^+_1$ transition employing 
the AMD(fix-s) calculation, which yields the $E4$ transition strength comparable to the 
empirical value of the $(e,e')$ data.
Noteworthily, the evaluation of $B(IS)/4$ by the $(\alpha,\alpha')$ data 
was inconsistent with the value, $B(C4)$, of the $(e,e')$ data, because the DWBA calculation 
was performed in the reaction analysis of 120~MeV $\alpha$-scattering data 
in Ref.~\cite{VanDerBorg:1981qiu}, although 
it was not applicable to the $4^+_1$ cross sections, as would be discussed subsequently.

Regarding the $E4$ transitions,  
we determine $f^\textrm{tr}$ by fitting the 
inelastic charge form factors to the $4^+_1$ and $4^+_2$ states measured by electron scattering.
The adopted $f^\textrm{tr}$ values and the transition strengths after the renormalization 
are listed in Table~\ref{tab:BEL}.
The values are $f^\textrm{tr}=1.4$ and 1.1 for the $4^+_1$ and $4^+_2$ states, respectively, 
implying that 40\% and 10\% enhancements are still required to fit the observed charge form factors. 

The squared elastic and inelastic charge form factors are shown in Fig.~\ref{fig:form}, which 
shows the comparison between the 
renormalized inelastic form factors and the data measured by electron scattering. 
The AMD(fix-s) calculation successfully describes the shapes of the experimental form factors, and 
precisely reproduces the shapes and magnitude of the observed form factors
after the renormalization.
The significant state dependence of the $E4$ transitions can 
be observed between the $4^+_1$ and $4^+_2$ states
in both the magnitude and shape of the form factors. 
In the form factors in the range of the transfer momentum, $q\le 3$~fm$^{-1}$, 
the $4^+_1$ state exhibits a narrow two-peak structure that is different from the broader shape of the $4^+_2$ form factors.
This result is the first microscopic calculation that successfully describes this unique character 
of the shape of the $4^+_1$ form factors, which had been a challenge to reproduce by structure calculations.

The calculated matter $\rho(r)$ and the renormalized $\rho^\textrm{tr}(r)$ 
of the AMD(fix-s) result are shown in Fig~\ref{fig:trans}. 
The unusual behavior of the $0^+_1\to 4^+_1$ transition density that corresponds to the 
two-peak structure of the charge form factors can be observed. 
For comparison, the Fermi density, 
\begin{align}
\rho^{}_\textrm{Fermi}(r)=\frac{\rho _0}{1+\exp(\frac{r-c}{t/4.4})},
\end{align}
with $c=2.876$~fm and $t=2.333$~fm, and the collective-model transition density of the Fermi-type Tassie form,
$\rho^\textrm{tr}_\textrm{Tassie}(r)\propto r^{\lambda-1}\partial \rho^{}_\textrm{Fermi}(r)/\partial r$,
are shown in the figure. $\rho^\textrm{tr}(r)$ of $0^+_1\to 4^+_2$ has the dominant peak
 at the nuclear surface, 
consistently with the standard $E4$ transition as shown by the collective-model $\rho^\textrm{tr}(r)$. 
Conversely, the behavior of $\rho^\textrm{tr}(r)$ of $0^+_1\to 4^+_1$ is much different from that for the standard $E4$ transition;
it exhibits an enhanced amplitude in the outer region and some suppression in the inner region. 
In the AMD(fix-s) result, this dominant amplitude in the outer region is produced by dynamical effects 
beyond the static mean-field, and its origin is determined by deconstructive contributions from various
types of excitations, including the $\alpha$-cluster dynamics at the nuclear surface and the $^{16}$O-core excitation.

\begin{figure*}[!thp]
\includegraphics[width=18. cm]{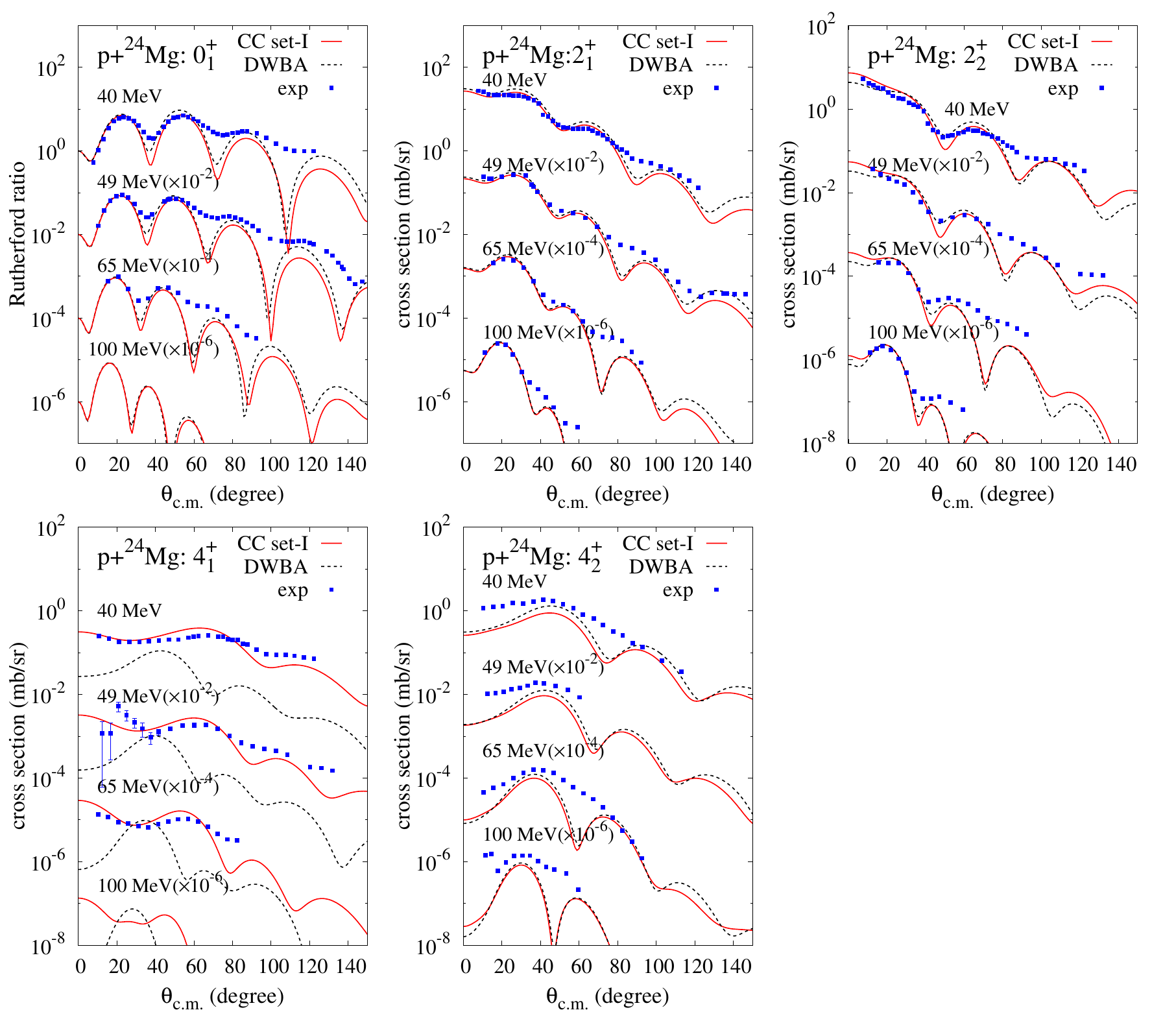}
  \caption{
Cross sections of the elastic and inelastic proton scattering off $\Mg$ 
at $E_p=40$, 49, 65, and 100 MeV calculated by
MCC+AMD (solid lines with label ``CC set-I'') and DWBA~(dotted lines) employing 
$f^\textrm{tr}$(set-I).
The experimental data are the cross sections at 
$E_p=40$~\cite{Zwieglinski:1978zza,Zwieglinski:1978zz},  49~\cite{Rush:1967zwr},
65~\cite{Kato:1985zz,EXFOR}, and 100~MeV~\cite{Horowitz:1969eso,EXFOR}.
  \label{fig:cross-mg24p}}
\end{figure*}

\begin{figure*}[!thp]
\includegraphics[width=18. cm]{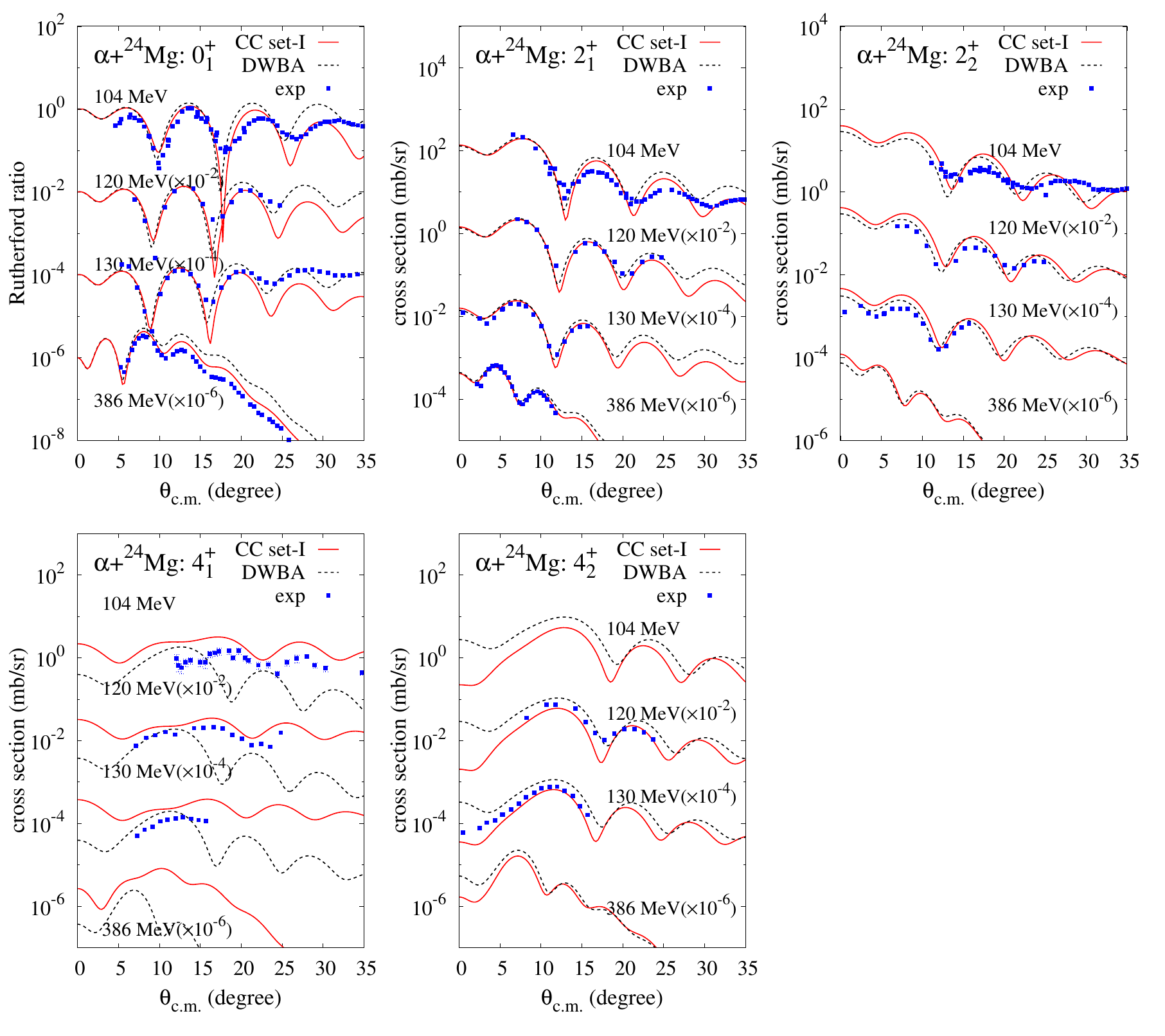}
  \caption{ 
Cross sections of elastic and inelastic $\alpha$-scattering off $\Mg$ at $E_\alpha=104$, 120, 130, and 
386 MeV calculated by
MCC+AMD (solid lines with label ``CC set-I'') and DWBA~(dotted lines) employing $f^\textrm{tr}$(set-I).
The experimental data are the cross sections 
at $E_\alpha=$104~\cite{Rebel:1972nip,EXFOR}, 120~\cite{VanDerBorg:1981qiu}, 130~\cite{Adachi:2018pql}, and 
386 MeV \cite{Adachi:2018pql}. 
  \label{fig:cross-mg24a}}
\end{figure*}

\begin{figure}[!thp]
\includegraphics[width=7. cm]{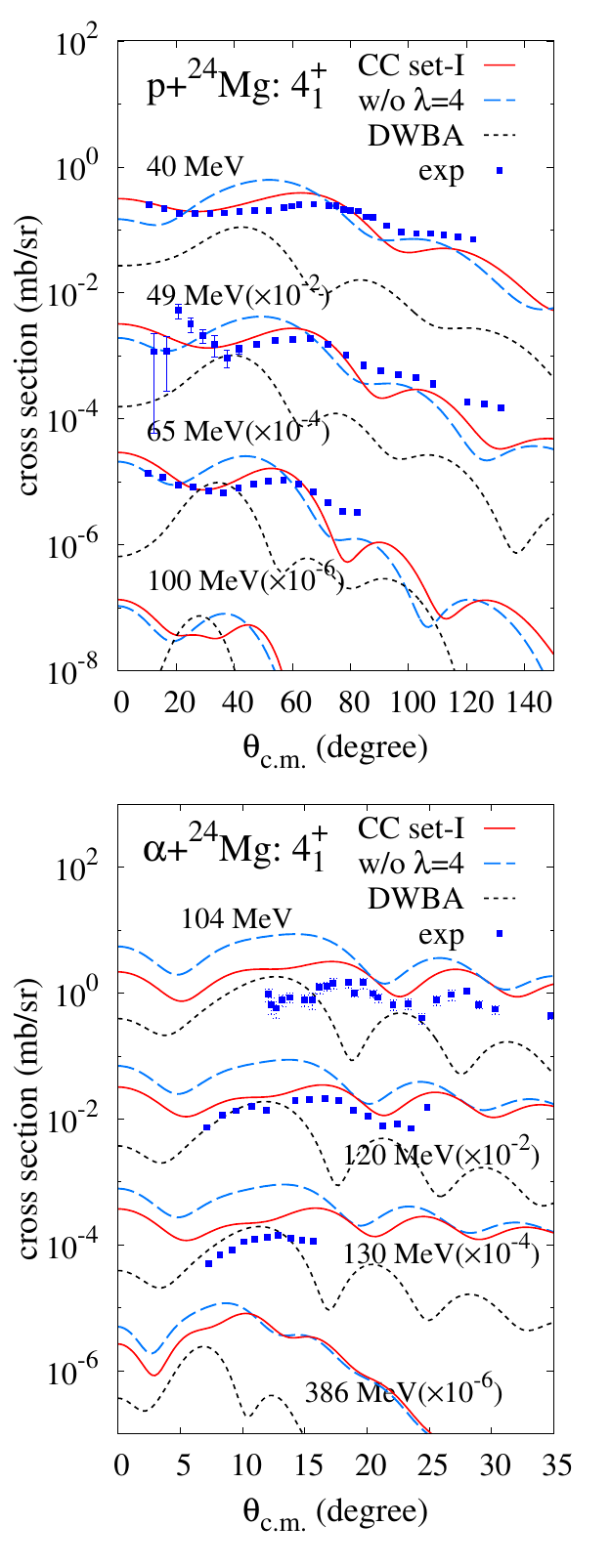}
  \caption{The MCC+AMD results with and without $\lambda=4$ transitions 
of the $(p,p')$ and $(\alpha,\alpha')$ cross sections to the $4^+_1$ state of $\Mg$, 
obtained with $f^\textrm{tr}$(set-I).
The full CC calculation with the $\lambda=4$ transitions~(CC set-I) 
and the CC calculation without the $\lambda=4$ transitions (w/o $\lambda=4$) are shown by solid
and dashed lines, respectively.
The one-step (DWBA) cross sections are shown by dotted lines. 
The experimental $(p,p')$ data are from Refs.~\cite{Zwieglinski:1978zza,Rush:1967zwr,Kato:1985zz,Horowitz:1969eso,EXFOR} and 
the $(\alpha,\alpha')$ data are from Refs.~\cite{Rebel:1972nip,EXFOR,VanDerBorg:1981qiu,Adachi:2018pql}.
  \label{fig:cross-mg24-m3}}
\end{figure}
\begin{figure}[!thp]
\includegraphics[width=7. cm]{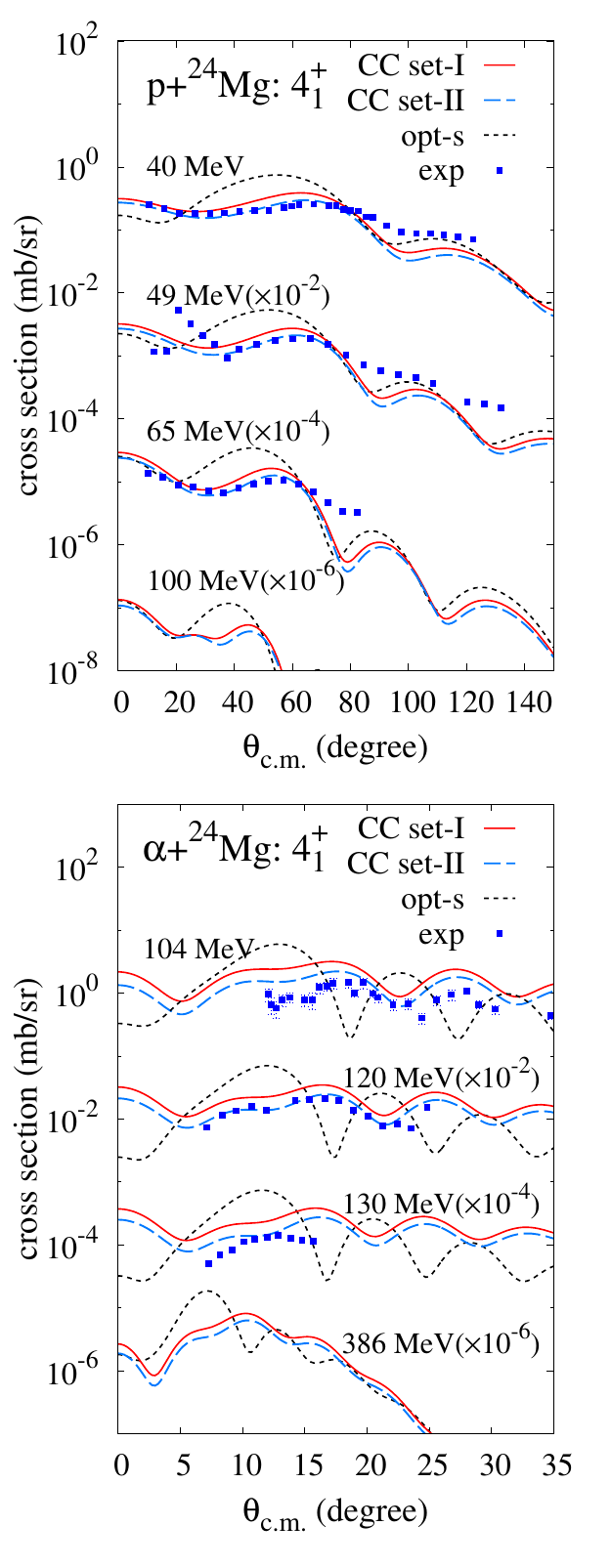}
\caption{Inelastic proton and $\alpha$-scattering cross sections 
of the $4^+_1$ state of $\Mg$ 
calculated by MCC+AMD employing the AMD(fix-s) densities in two cases of the renormalization factors,
$f^\textrm{tr}$(set-I) and $f^\textrm{tr}$(set-II). 
The previous MCC+AMD results employing the AMD(opt-s) densities~\cite{Kanada-Enyo-mg24} 
are also shown. 
The experimental $(p,p')$ data are taken from Refs.~\cite{Zwieglinski:1978zza,Rush:1967zwr,Kato:1985zz,Horowitz:1969eso,EXFOR} and 
the $(\alpha,\alpha')$ data are from Refs.~\cite{Rebel:1972nip,EXFOR,VanDerBorg:1981qiu,Adachi:2018pql}.
  \label{fig:cross-mg24-m2}}
\end{figure}

\section{Proton and $\alpha$-scattering}  \label{sec:results2}

MCC+AMD calculations are performed for the proton and $\alpha$-scattering 
employing the diagonal and transition densities obtained by AMD(fix-s). 
The $J^\pi=0^+_1$, $2^+_1$, $2^+_2$, $4^+_1$, and $4^+_{2}$
states of $\Mg$ and $\lambda=0$, 2, and 4 transitions between them are included
in the CC calculations, which we call the full CC calculations.
In the default MCC+AMD calculations, the calculated transition densities are renormalized by 
the factors, $f^\textrm{tr}$(set-I), which are listed in Tables~\ref{tab:BE2} and \ref{tab:BEL}. 
We calculate the elastic and inelastic cross sections of the  
proton scattering at incident energies of $E_p=40$, $49$, 65, and $100$~MeV and 
$\alpha$-scattering at $E_\alpha=104$, 120, 130, and 386~MeV. 
To observe the CC effects, the one-step cross sections are also calculated by
DWBA employing the same inputs.
It should be commented that 
procedures of the reaction calculations in this paper are, in principle, the same as those
in the previous paper~\cite{Kanada-Enyo-mg24} except 
for the structure inputs; the revised diagonal and transition densities of $\Mg$ 
obtained by AMD(fix-s) are utilized in the present calculation instead of the 
AMD(opt-s) densities in the previous calculation.

The calculated proton scattering cross sections are shown  
in Fig.~\ref{fig:cross-mg24p} and compared to the experimental data. 
The present MCC+AMD calculation reasonably reproduces the amplitudes 
of the elastic and inelastic cross sections of proton scattering within this energy range. 
It also reproduces the diffraction patterns around the peak positions. Notably,
the dip structure in the MCC+AMD results should be smeared by spin-orbit potentials, which are ignored 
in this calculation.
Compared to the DWBA calculation, it is observed that the one-step process contributes
dominantly to the proton scattering cross sections except in the $4^+_1$ state. 
For the proton inelastic scattering to the $4^+_1$ state, the MCC+AMD calculation reproduces
the amplitude and angular distribution around the first and second peaks whereas 
the DWBA cross sections are inconsistent in both amplitude and angular distribution. 
The successful result for the $4^+_1$ cross sections, which could not be obtained 
by the  AMD(opt-s) densities in the previous 
MCC+AMD calculation, is a new result of the 
the present MCC+AMD calculation by employing the AMD(fix-s) densities.

The results of the $\alpha$-scattering are shown in Fig.~\ref{fig:cross-mg24a}.
The calculated elastic and inelastic cross sections are compared to the experimental data. 
The MCC+AMD calculation adequately reproduces the amplitudes and angular
distributions of the existing data at an energy range of $E_\alpha=100$--400~MeV.
The good agreements with the experimental cross sections for the  $0^+_1$, $2^+_1$, $2^+_2$, and $4^+_2$ states
are obtained by the present MCC+AMD calculation employing the AMD(fix-s) densities
consistently 
with the previous results obtained with the AMD(opt-s) densities. 
The present calculation also succeeds in reproducing the $4^+_1$ cross sections,
for which the AMD(opt-s) densities in the previous 
MCC+AMD calculation failed to describe the experimental data.
Compared to the one-step cross sections of DWBA,
significant CC effects are observed in the $4^+_1$ and $4^+_2$ cross sections. 
Particularly, the one-step cross sections of the $4^+_1$ state are inconsistent with 
the amplitude and angular distribution of the observed 
$(\alpha,\alpha')$ data, thereby implying that the multi-step process is also essential 
for the $\alpha$ inelastic scattering to the $4^+_1$ state.

To acquire more details on the direct and multi-step contributions to the $4^+_1$ cross sections 
of proton and $\alpha$-scatterings,
we conduct the CC calculations by including the $\lambda=0,2$ transitions and omitting 
the $\lambda=4$ transitions. We compare the results to the full CC result including 
the $\lambda=0,2,4$ transitions. 
Figure~\ref{fig:cross-mg24-m3} displays the results with and without the $\lambda=4$ transitions for the 
$4^+_1$ cross sections of proton and $\alpha$-scatterings together with the one-step cross sections of DWBA.
In the CC results without the $\lambda=4$ transitions, 
the two-step process via the $E2;0^+_1\to 2^+_1$ and $E2;2^+_1\to 4^+_1$ transitions 
delivers the dominant contribution 
to the $4^+_1$ cross sections of proton and $\alpha$-scattering. Compared to the two-step contribution, 
the one-step contribution is more than one-order weaker. However, in the full CC results,  
the one-step process suppresses the dominant two-step contribution, in particular, 
in the $\alpha$-scattering cross sections 
and significantly alters the angular distribution of the $4^+_1$ cross sections. 
This deconstructive interference of the dominant two-step contribution
by the one-step process is essential in describing the observed diffraction patterns. 

Notably, in the energy systematics, the observed 49~MeV $(p,p')$ data at the forward angles 
were inconsistent with the data at $E_p=40$ and 65~MeV in the other experiments. 
Additionally, in the $\alpha$-scattering data, 
the $130$~MeV $(\alpha,\alpha')$ data were inconsistent with the data at $E_\alpha=104$ and 120~MeV.
Those data should be reexamined by the careful separation of 
the weak $4^+_1$(4.12~MeV) spectra from the strong $2^+_2$ spectra of the inelastic scatterings. 

Figure~\ref{fig:cross-mg24-m2} compares
the present and previous results of  MCC+AMD for the $4^+_1$ cross sections 
obtained with the AMD(fix-s) and AMD(opt-s) densities, respectively.
The present calculation employing the AMD(fix-s) densities successfully reproduces the angular distributions 
of the $(p,p')$ and $(\alpha,\alpha')$ cross sections, 
whereas inconsistent results were obtained in the previous calculation using the AMD(opt-s) densities
because the $E4;0^+_1\to 4^+_1$ transition density of AMD(opt-s) was incorrect 
(refer to the charge form factors in Fig.~\ref{fig:form}(d)). 
In other words, 
the success of the present MCC+AMD calculation is because of
the detailed description of the $0^+_1\to 4^+_1$ transition density, which is essential for 
the deconstructive interference of the strong two-step process by
the weak one-step process in the $4^+_1$ cross sections.
This result indicates that the MCC approach is beneficial
to examine the validity of the structure inputs even if the one-step contribution is not dominant. 
Noteworthily, it is proved that the MCC approach combined with the  
microscopic structure calculation is necessary to solve the puzzle in the $4^+_1$ cross sections. 

In the detailed comparison of the preset result with the observe data, 
there is a slight overestimation of the $(\alpha,\alpha')$ cross sections of the $4^+_1$ state.
To observe the effect of the ambiguity in the $E2;4^+_1\to 2^+_1$ transition strength to the $4^+_1$ cross sections, 
we conduct the MCC+AMD calculation for $f^\textrm{tr}$(set-II). 
Figure~\ref{fig:cross-mg24-m2} shows the comparison of the results of $f^\textrm{tr}$(set-I) 
and $f^\textrm{tr}$(set-II).
The difference between sets I and II calculations is only a slight modification in the 
renormalization factor of the $E2;4^+_1\to 2^+_1$ transition density; it changes
from $f^\textrm{tr}$(set-I)=1.26 
to $f^\textrm{tr}$(set-II)=1.11. Regarding set-II, $f^\textrm{tr}$(set-II)
of 1.11 for $4^+_1\to 2^+_1$ coincides with 
$f^\textrm{tr}$ of 1.12 for the $2^+_1\to 0^+_1$ transition in the same band. 
By this modification, 
the $4^+_1$ cross sections are slightly reduced 
especially for the $\alpha$-scattering, and better results are obtained
in set II than in set I.
This implies that the value, $f^\textrm{tr}=1.11$, for 
the $E2;4^+_1\to 2^+_1$ transition, which is adjusted to $B(E2;4^+_1\to 2^+_1)=125$~$e^2\textrm{fm}^4$,  
is favorable to describe the $4^+_1$ cross sections in the present MCC+AMD approach, 
although we can not propound a definite conclusion because of the possible ambiguities in other transitions.

\section{Summary} \label{sec:summary}
The structure and transition properties of the $K^\pi=0^+$ and $K^\pi=2^+$ bands of $^{24}$Mg were investigated 
by the microscopic structure and reaction calculations via inelastic proton and 
$\alpha$-scattering. 
For the structure calculation, 
we adopted AMD+VAP with fixed nucleon-spins called AMD(fix-s),
instead of the previous AMD+VAP calculation with optimized nucleon-spins, which we called AMD(opt-s).
Employing the revised structure calculation~(AMD(fix-s)), 
the $E4$ transitions to the $4^+_1$ and $4^+_2$ states were reexamined.

Employing the AMD(fix-s) calculation of the $\Mg$ structure,  
improved results of the energy of the $K^\pi=2^+$ band and 
the $E2$ and $E4$ transition strengths, which 
the previous AMD(opt-s) calculation could not quantitatively reproduce, were obtained. 
These improvements were obtained because of the enhanced triaxiality achieved 
by the AMD(fix-s) calculation. 
For the transition properties of the $4^+$ states,
a significant improvement was obtained for the $E4$ charge form factors by the AMD(fix-s) result. 
The remarkable difference between the $4^+_1$ and $4^+_2$ states of the charge form factors, 
which was observed by electron scattering, was described well by the AMD(fix-s) calculation. 
Particularly, the calculation successfully reproduced
the unusual behavior of the $4^+_1$ form factors with a narrow two-peak structure. 
This structure corresponded to the enhanced outer amplitude of $\rho^\textrm{tr}$, 
which was different from the conventional shape of standard $E4$ transitions. 
This is the first microscopic calculation that described this unique property of 
the $E4;0^+_1\to 4^+_1$ transition. 

Employing the renormalized AMD(fix-s) densities, 
the MCC calculations of the proton and $\alpha$-scattering off $\Mg$ were performed 
utilizing the Melbourne $g$-matrix effective interaction.
The MCC+AMD calculation in this study reasonably reproduced 
the elastic and inelastic cross sections of proton scattering in the energy range of $E_p=40$--100~MeV.
For the $\alpha$-scattering, the calculation reproduced the amplitudes and angular
distributions of the existing data adequately at an energy range of $E_\alpha=100$--400~MeV.
Particularly, the present calculation with the AMD(fix-s) densities was in good
good agreement with the 
$(p,p')$ and $(\alpha,\alpha')$ data for 
the $4^+_1$ state, which has long been a challenge to explain via reaction calculations. 
In the detailed analysis of the CC calculations with and without the $\lambda=4$ transitions 
and DWBA calculation,
it was revealed that the dominant two-step process via the $E2;0^+_1\to 2^+_1$ and $E2;2^+_1\to 4^+_1$ transitions and the 
deconstructive interference by the weak one-step process were essential for the inelastic 
proton and $\alpha$-scattering to the $4^+_1$ state. 

The present results indicated that the MCC approach was beneficial 
for examining the validity of the structure inputs even if the one-step contribution 
was not dominant. 
The MCC approach combined with the microscopic structure calculation, which afforded correct charge form factors, was necessary
for solving the puzzle in the proton and $\alpha$ inelastic scattering to the $4^+_1$ state of $\Mg$.

\begin{acknowledgments}
The computational calculations of this work were performed using the
supercomputer at the Yukawa Institute for Theoretical Physics at Kyoto University. The work was supported
by Grants-in-Aid of the Japan Society for the Promotion of Science (Grant Nos. JP18K03617, JP16K05352, and 18H05407) and by a grant of the joint research project of the Research Center for Nuclear Physics at Osaka University.
\end{acknowledgments}

\end{document}